\documentclass[%
,preprint%
,secnumarabic%
,amssymb, amsmath,nobibnotes, aps, prl]{revtex4}

\usepackage{graphicx}
\usepackage[usenames]{color}

\begin{document}

\title{Halving the Casimir force with conductive oxides}

\author{S.~de Man}
\author{K.~Heeck}
\author{R.~J.~Wijngaarden}
\author{D.~Iannuzzi}
\email{iannuzzi@few.vu.nl}
\affiliation{Department of Physics and Astronomy, VU University
Amsterdam, De Boelelaan 1081, 1081 HV Amsterdam, The Netherlands}

\date{\today}

\begin{abstract}
The possibility to modify the strength of the Casimir effect by tailoring the dielectric functions of the interacting surfaces is regarded as a unique opportunity in the development of Micro- and NanoElectroMechanical Systems. In air, however, one expects that, unless noble metals are used, the electrostatic force arising from trapped charges overcomes the Casimir attraction, leaving no room for exploitation of Casimir force engineering at ambient conditions. Here we show that, in the presence of a conductive oxide, the Casimir force can be the dominant interaction even in air, and that the use of conductive oxides allows one to reduce the Casimir force up to a factor of 2 when compared to noble metals.
\end{abstract}

\maketitle

The mechanical parts of Micro- and NanoElectroMechanical Systems (MEMS and NEMS) are often designed
to work at separations where the Casimir effect \cite{casimir} might play a relevant role \cite{vdwmems,chanscience,channonlinear,memsstiction}. It is thus commonly believed that, if one could suitably engineer the strength of the Casimir force, unprecedented
opportunities would come available for the development of conceptually new MEMS and NEMS \cite{ieeecapasso,repulsivevacuum,torque,capassonature2008}. The most simple approach to tailor the Casimir force is to properly choose the materials of which the interacting surfaces are made. According to the Lifshitz theory \cite{lifshitz}, the interaction between two objects depends on their dielectric functions.
Transparent dielectrics, for example, attract less than reflective mirrors. This property may be used to reduce the Casimir attraction whenever the design requires a smaller short range interaction. It is however fair to say that, for the vast majority of applications, MEMS and NEMS operate in air, where surfaces tend to accumulate trapped charges. Those charges give rise to a strong electrostatic interaction that cannot be compensated by a counterbias voltage and that typically overcomes the Casimir force. It is thus difficult to imagine that the Casimir force can play an important role in MEMS and NEMS operating in air, unless all surfaces are coated with noble metals to reduce the forces due to surface charges to negligible levels. In that case, however, there is not much room to tune the strength of the Casimir interaction because the diversity in the dielectric functions of different metals is simply not large enough
\cite{deccaCu,pnas,light,skin,mohideenSi}. As a matter of fact, to date, there is no experiment that shows that, in air, the Casimir force can still be tuned significantly while remaining the dominant interaction mechanism.

In this paper we present a precise measurement of the Casimir force between a gold coated sphere
and a glass plate coated with either a thick gold layer or a highly conductive, transparent oxide
film. The experiment was performed in air, and no electrostatic force due to residual trapped charges was
observed over several weeks of measurements in either case. The decrease of the Casimir force due
to the different dielectric properties of the reflective gold layer and the transparent oxide film
resulted to be as high as $\simeq 40\% - 50\%$ at all separations (from $\simeq 50$ to $\simeq
150$~nm). Our experiment shows that, in the presence of a conductive oxide layer, the Casimir force can still be the dominant interaction mechanism even in air, and indicates that, whenever the design might require it, it is possible to tune the Casimir attraction by a factor of 2.

\begin{figure}[h!]
\includegraphics[width=13cm]{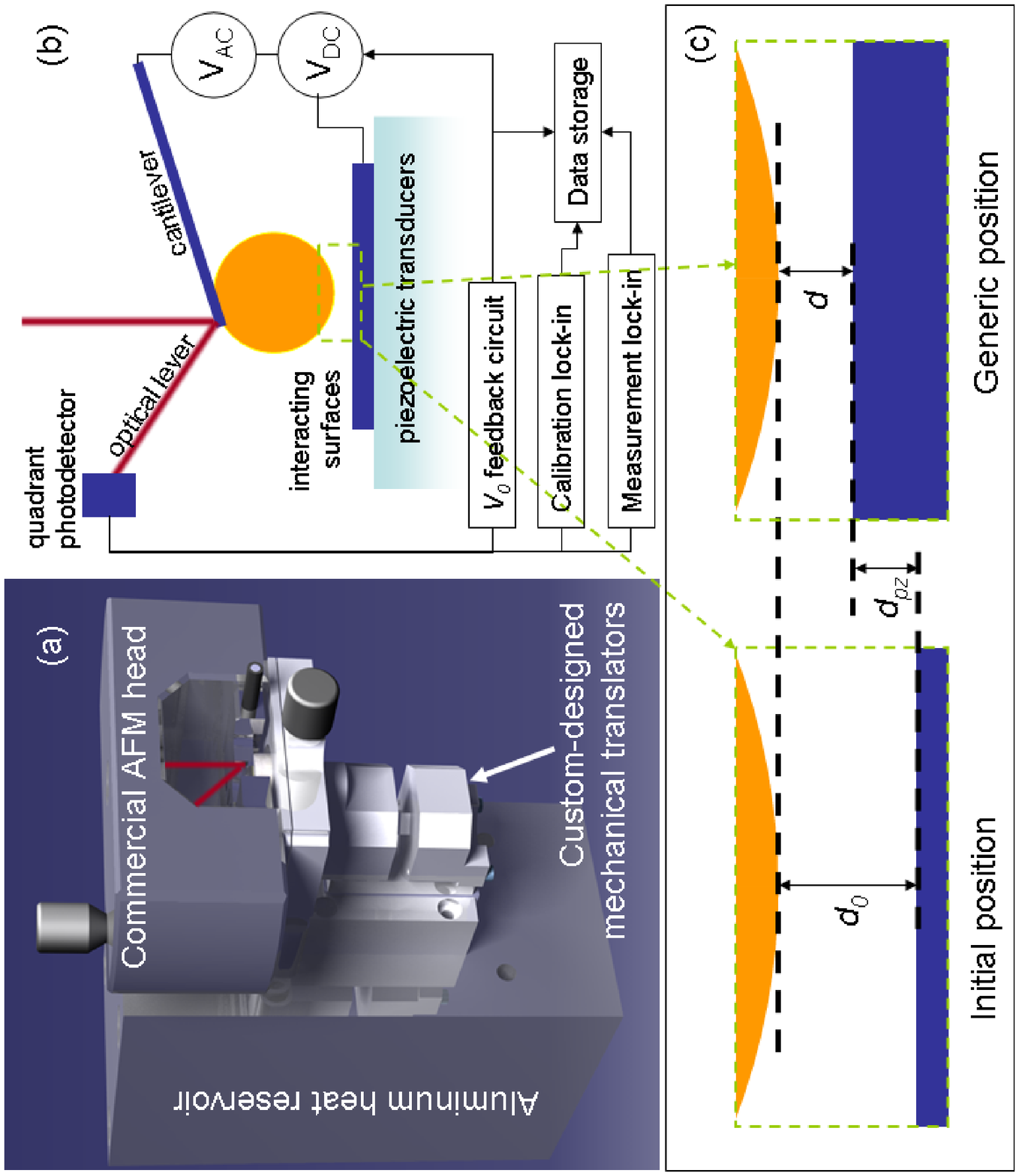}
\caption{} \label{fig:schematic}
\end{figure}

Our experimental set-up is designed to perform precise measurements of surface forces between a 100
$\mu$m radius sphere and a plate as a function of their separation (see Fig. \ref{fig:schematic} and
\cite{pra}). The sphere is glued onto the hanging end of a micromachined cantilever (spring constant $\approx 1$ N/m, resonance frequency (with the sphere attached) $\approx 1.9$ kHz). The plate is
mounted on a capacitive feedback controlled piezoelectric transducer that allows one to accurately
vary the separation between the sphere and the plate in discrete steps. Any force acting between
the two surfaces results in a bending of the cantilever that is detected by the optical lever of a
commercial Atomic Force Microscope (AFM) head \cite{afm}. The set-up is kept at a fixed temperature
to reduce mechanical drifts and is placed on an active anti-vibration stage inside an acoustic
isolation box to decouple the force sensor from external vibrations.

Precise Casimir force measurements typically require careful analysis of three crucial issues.
First, even when both the sphere and the plate are coated with metallic films, there might still
exist an electrostatic potential difference $V_0$ between the two surfaces that gives rise to a
residual electrostatic force. This force must be actively canceled by counter biasing $V_0$ with an
externally applied voltage. Second, although the relative displacements of the piezoelectric
transducer that moves the plate, $d_{pz}$, are precisely controlled, the separation between the two
interacting surfaces at the start of the measurement, $d_0$, is \textit{a priori} unknown (see Fig.
\ref{fig:schematic}). Therefore, the absolute separation $d=d_0-d_{pz}$ has to be obtained from a
calibration procedure. Third, the electronic signal coming out of the AFM head must be converted
into force. It is thus necessary to calibrate the instrument with a controlled force. To address
these problems, we have designed a measurement technique that allows one to simultaneously: (i)
compensate for the residual potential, (ii) calibrate the set-up, and (iii) measure the Casimir
force \cite{pra,inprep}. In a nutshell, in each calibration/measurement run the plate is moved in
discrete steps from $d_0$ to a minimum value of $d$ (just before contact with the sphere). A
calibrated AC electrostatic potential is applied between the sphere and the plate at a frequency
$\omega_1$ much lower than the resonance frequency of the force sensor ($\frac{\omega_1}{2\pi}=72.2$ Hz). This AC excitation is used
to drive a feedback circuit that compensates for the residual voltage $V_0$, and, simultaneously,
generates an electrostatic force that makes the cantilever oscillate at $2\omega_1$. The amplitude
of the oscillations, measured with a lock-in amplifier (calibration lock-in in
Fig.~\ref{fig:schematic}b), are recorded as a function of $d_{pz}$, and are then used to calibrate
the instrument and extract $d_0$. At the same time, a transducer mechanically coupled to the
piezoelectric translator makes the plate move around $d_{pz}$ with an amplitude of $3.85 \pm
0.08$~nm at a frequency $\omega_2$, which is again much lower than the resonance frequency of the
force sensor ($\frac{\omega_2}{2\pi}=119$ Hz) \cite{carugnoruoso}. In the presence of a force that depends on separation (e.g., the
Casimir force), the cantilever bends in phase with the modulation of $d$. The amplitude of the
in-phase oscillation, measured with another lock-in amplifier (measurement lock-in in
Fig.~\ref{fig:schematic}b), is proportional to the derivative of the force with respect to $d$.
Furthermore, the presence of the cushion of air between the two surfaces gives rise to a
hydrodynamic force that depends on the velocity with which the plate moves. The signal produced by
this force manifests itself at the same frequency at which $d$ is modulated, but with a phase
rotated by 90~degrees. This contribution does not influence the output of the in-phase component
and can be measured independently with the same lock-in amplifier. The integration times of the
lock-in amplifiers are 8~s for every value of $d_{pz}$. A typical measurement run consists of
$\simeq 50$ $d_{pz}$ set-points in the measurement range $50 < d < 1100$~nm, and takes roughly 7
minutes. The cantilever responses to the modulations at $\omega_1$ and $\omega_2$ are $<80$~pm
(root-mean-square) during the entire experiment. All force measurements are performed in air at
atmospheric pressure, temperature 300~K, and 29\% relative humidity.

In this paper we present two experiments performed with the same sphere and two different plates.
The sphere is a polystyrene sphere with nominal radius 100~$\mu$m coated with a Ti adhesion layer
followed by a 100~nm Au film (surface roughness 3.8~nm RMS). The plate used in the first
experiment is a polished sapphire substrate coated with a metallic film similar to the one
deposited on the sphere (surface rougness 0.8~nm RMS). The plate used in the second
experiment is a float glass substrate with a 190 nm Indium Tin Oxide (ITO, In$_2$O$_3$:Sn)
sputtered thin film on top (PGO CEC010S, typically $8.5~\Omega/\Box$, or, equivalently, $\rho =
1.6~10^{-4}~\Omega$cm, total surface roughness 4~nm RMS). We have measured the reflection and
transmission spectra of both plates in the wavelength range $180~\mathrm{nm} < \lambda < 2.5~\mu$m,
and observed that the optical properties of our films are in agreement with the literature
\cite{palik,fujiwara}.

\begin{figure}[h!]
\includegraphics[width=13cm]{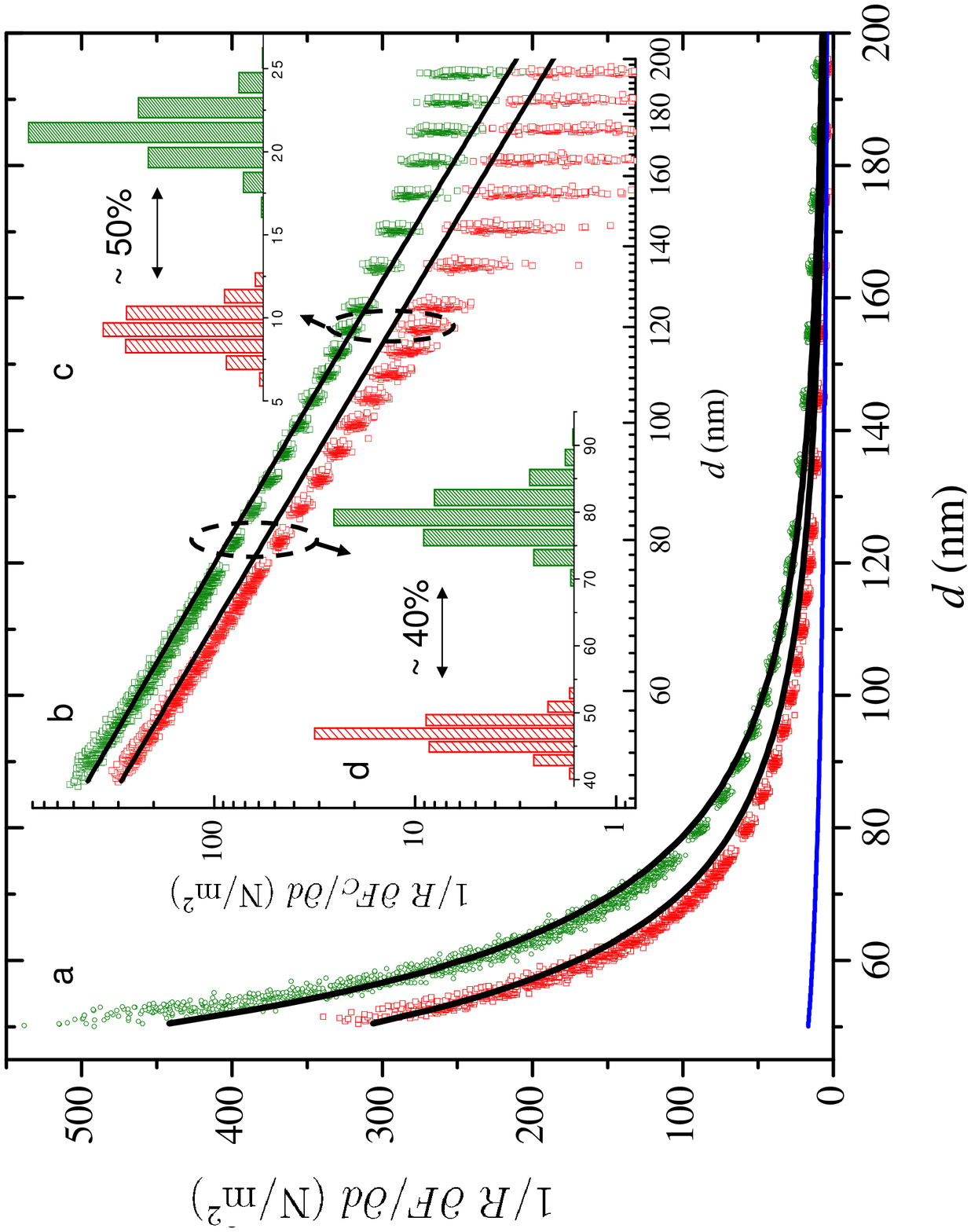}
\caption{} \label{fig:forceresults}
\end{figure}

Fig.~\ref{fig:forceresults}a presents measurements of the force between the sphere and the
plates coated with either Au or ITO \cite{noteond0}. The experimental data represent the spatial
derivative of the total force (normalized by the sphere radius $R$), which is the sum of the
Casimir interaction, a Coulomb interaction induced by the presence of trapped charges (if any), and an electrostatic attraction due to the AC calibration potential. The
strength of the latter can be estimated from the simultaneous calibration procedure \cite{inprep}.
From Fig.~\ref{fig:forceresults}, it is evident that this electrostatic contribution, which is
anyway equal in both experiments (within 2\%), is small compared to the total force signal. The
black lines in Fig.~\ref{fig:forceresults}a are computations of the Casimir force using the Lifshitz theory \cite{parsegian} with dielectric functions calculated as in \cite{fujiwara} (for ITO) and \cite{deman} (for Au); the electrostatic force due to the calibration potential
is added to the theory in order to compare with the raw data. The calculation of the Casimir force should only be
considered approximate, because the dielectric functions of the samples are not known precisely
\cite{svetovoy}, and no surface roughness corrections are applied \cite{noteond0}. Still, the agreement between the calculation and the data shows that the Casimir effect largely dominates any Coulomb interaction that would have been otherwise observed in the presence of a significant amount of trapped charges.
Fig.~\ref{fig:forceresults}b shows the data and theory on a double logarithmic scale, where we have
subtracted the electrostatic background due to the AC calibration potential using the simultaneously obtained calibration
data. At small separations $d<60$~nm, both data sets curve upwards, which might be a sign of
surface roughness effects. At separations $d>120$~nm, the experimental data for the Au-Au Casimir
interaction start to deviate significantly from the theory because of an artefact caused by
reflections from the optical lever light by the sample. This reflected light reaches the
photodetector and causes a background signal that is not related to any force. This artefact is
common to all optical lever based AFM techniques, and the related signal is typically assumed to be
linear in the piezo extension and subtracted from the data accordingly \cite{mohideenPRLold}.
Because the reflectivities of our two samples are so different, we prefer to refrain from such a
procedure, and present the data as is. In Figs.~\ref{fig:forceresults}c and d we show histograms of
all the obtained measurements for the derivative of the Casimir force for two specific separations
($d=120$~nm and $d=80$~nm). The histograms at $d=80$~nm can be described by Gaussians with a
standard deviation of roughly 5\%, which means that our method provides a precision in the mean
measured Casimir force derivative of 0.2\%. It is evident from the histograms that the spatial
derivative of the Casimir force between a Au and an ITO surface is roughly $\simeq 40\% - 50\%$
smaller than between two Au surfaces. In our geometry (i.e., for separations much smaller than the
radius of the sphere), the spatial derivative of the force is proportional to the pressure between
two parallel plates \cite{parsegian}. We can thus conclude that the Casimir pressure that one would
measure between a Au plate kept parallel to an ITO plate would be roughly $\simeq 40\% - 50\%$
smaller than in the case of two Au plates.

Even though the agreement between the theoretical prediction and the measurement of the Casimir
force in both situations is good, one might still argue that the observed decrease could be
mimicked by drifts in $d_0$. For both measurements series, we gathered 580~data sets continuously,
which allows us to directly assess the run-to-run drift of $d_0$. Due to the temperature
stabilization of our setup, the mechanical drift is very small at $\simeq 0.1$~nm and $\simeq
0.2$~nm per run for the Au-Au and the Au-ITO experiments, respectively. We conclude that the
decrease of the Casimir force cannot be ascribed to drifts in $d_0$. We have also verified that the
electrostatic force used to calibrate the instrument and extract $d_0$ follows what expected from
elementary electrostatic arguments, as suggested in \cite{roberto} and discussed in \cite{pra}.
Concerning the compensation voltage, we observed that $V_0$ varies approximately $1$~mV and $3$~mV
over the complete measurement range in the Au-Au and Au-ITO cases, respectively \cite{note_on_v0}. These slight
variations of $V_0$ do not compromise the measurement of the Casimir force at the current level of
sensitivity.

Finally, the different surface roughnesses of the sphere and the plates also influence the strength
of the Casimir effect. Since both experiments are conducted with the same sphere, the difference in
the observed Casimir force can never be due to the surface roughness of the sphere. Second, we
recall that the surface roughness of the ITO sample is larger than that of the Au substrate. Since
surface roughness enhances the Casimir force \cite{maradudin} we note that, if it played a role in the
probed separation range, it would lead to a stronger interaction between Au and ITO than between
two Au surfaces, contrary to the measurements presented in Fig.~\ref{fig:forceresults}.

\begin{figure}[h!]
\includegraphics[width=13cm]{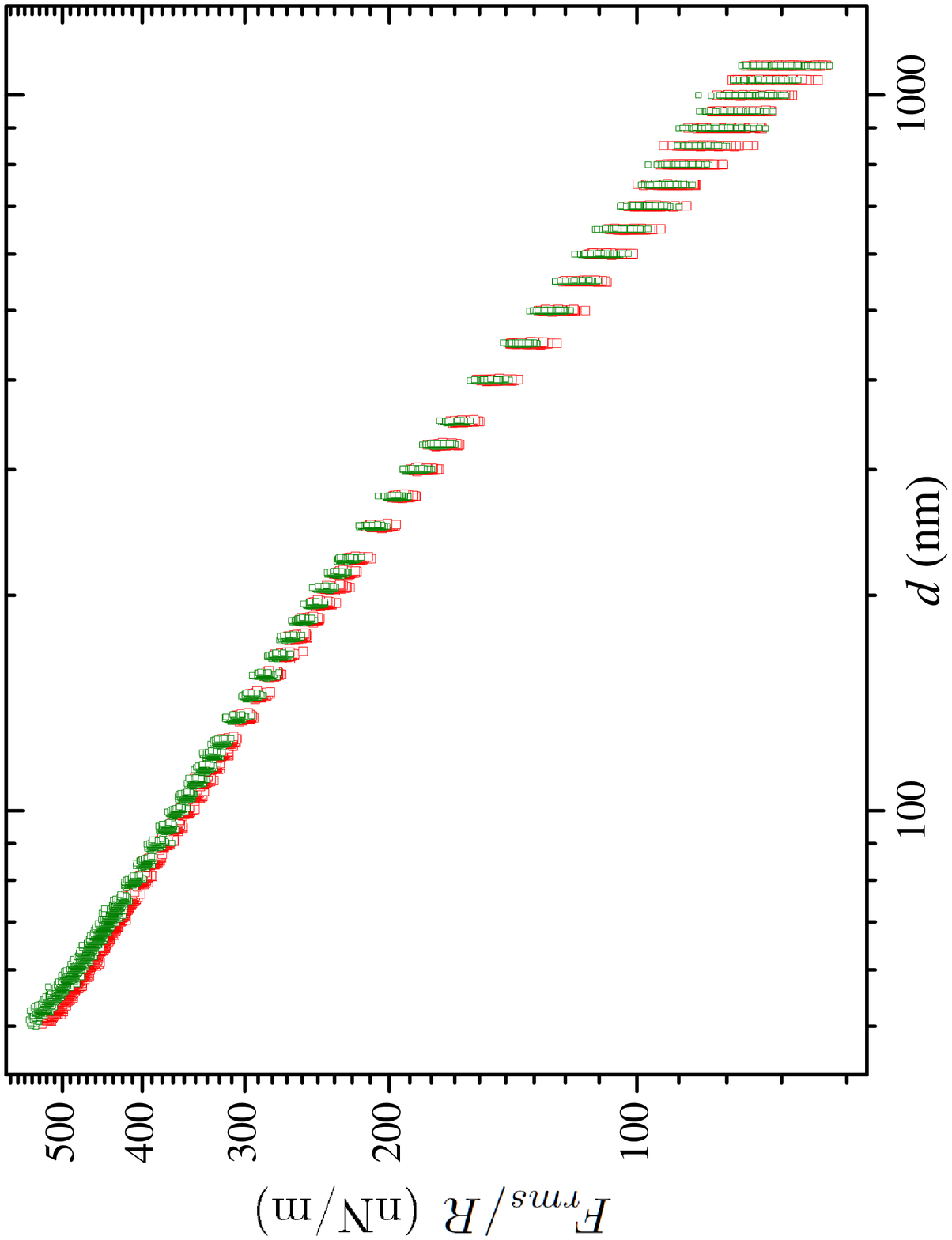}
\caption{} \label{fig:hydroresults}
\end{figure}

To make our claim even more robust, we can now compare the hydrodynamic force observed during the
two experiments. Because the geometrical configuration of the experiment is equal in both cases, we
expect to measure the same hydrodynamic force. The results are shown in
Fig.~\ref{fig:hydroresults}. It is clear that the hydrodynamic force is very similar in both cases,
although there exists a slight discrepancy between the two curves ($\simeq 2\%$). Since both curves
are parallel on a double logarithmic scale, we conclude that this discrepancy cannot be ascribed to
a difference in the calibration of $d_0$. Therefore, we rule out that the large difference in the
Casimir force reported in Fig.~\ref{fig:forceresults} be due to an error in the determination of
$d_0$.

In conclusion, we have demonstrated that an ITO coating of one of the two surfaces is sufficient to readily create situations where the Casimir force is still the dominant interaction mechanism regardless the presence of air in the surroundings. Since ITO is transparent over a wide range of frequencies, the Casimir attraction is up to a factor of 2 smaller when compared to the case of the Au-Au interaction, leaving ample room for Casimir force engineering even at ambient conditions, where MEMS and NEMS typically operate.

The authors thank A.~Baldi, F.~Mul, U.~Mohideen, J.~H.~Rector, B.~Dam, and R.~Griessen for useful
discussions. This work was supported by the Netherlands Organisation for Scientific Research (NWO),
under the Innovational Research Incentives Scheme VIDI-680-47-209. D.I. acknowledges financial
support from the European Research Council under the European Community's Seventh Framework
Programme (FP7/2007-2013)/ERC grant agreement 201739.

\newpage

\noindent FIG. 1. \textbf{a,} Drawing of the experimental set-up. \textbf{b,} Schematic
representation of the working principle of the experimental technique. \textbf{c,} Definition of
$d_0$ (initial separation), $d_{pz}$ (movement of the piezoelectric stage), and $d$ (separation
between the two surfaces).

\newpage

\noindent FIG. 2. \textbf{a,} Spatial derivative of the total force as a function of absolute
surface separation for the Au-Au (green squares) and Au-ITO (red squares) interactions for randomly
chosen subsets of the data (150 out of 580 for both cases). The blue line represents the derivative
of the electrostatic force caused by the simultaneous calibration procedure (common to both the
gold and ITO measurements). The black lines indicate the calculated Casimir forces with the
electrostatic background added. \textbf{b,} Spatial derivative of the Casimir force, with the
electrostatic background subtracted from the data. The black lines correspond to the calculations
of the Casimir force. \textbf{c,} Histograms of 580 force measurements for Au-Au and Au-ITO at
$d=120$~nm. \textbf{d,} Same as c, but for $d=80$~nm.

\newpage

\noindent FIG. 3. Hydrodynamic force acting on the sphere as a function of the absolute separation
$d$ for Au-Au (green squares) and Au-ITO (red squares), for the data sets shown in
Figs.~\ref{fig:forceresults}a and~b.

\newpage

\begin{thebibliography}{99}
\bibitem{mohideenSi} $\simeq 30\%$ variations of the Casimir force were observed in
U. Mohideen \textit{et al.}, Phys. Rev. A \textbf{72}, 020101(R) (2005). Also this strategy is not
practical for MEMS in air, because the passivation layer is not stable under atmospheric
conditions.
\end{thebibliography}
\end{document}